\documentstyle[prl,preprint,eqsecnum,aps,epsf]{revtex}

\begin{document}
\bibliographystyle{plain}
\title{Singular Quantum Mechanical Viewpoint of Localized Gravity
in Brane-World Scenario}
\author{D. K. Park and Hungsoo Kim}
\address{Department of Physics, Kyungnam University,
Masan, 631-701, Korea.}
\date{\today}
\maketitle

\begin{abstract}
The graviton localized on the $3$-brane is examined in Randall-Sundrum
brane-world scenario from the viewpoint of one-dimensional singular
quantum mechanics. For the Randall-Sundrum 
single brane scenario the one-parameter family 
of the fixed-energy amplitude is explicitly computed where the free 
parameter $\xi$ parametrizes the various boundary conditions at the brane.
The general criterion for the localized graviton to be massless is derived 
when $\xi$ is arbitrary but non-zero. When $\xi=0$, the massless graviton 
is obtained via a coupling constant renormalization. For the two branes 
picture the fixed-energy amplitude is in general dependent on the two
free parameters. The numerical test indicates that there is no massless
graviton in this picture. For the positive-tension brane, however, the 
localized graviton becomes massless when the distance between branes are
infinitely large, which is essentially identical to the single brane
picture. For the negative-tension brane there is no massless graviton regardless
of the distance between branes and choice of boundary conditions.
\end{abstract}
\newpage
\section{Introduction}
The first Randall-Sundrum(RS1) brane-world scenario\cite{rs99-1} was 
designed to solve the
gauge hierarchy problem which is one of the longstanding puzzle in
physics. To examine this problem they have introduced two branes located at the 
boundary of the compactified fifth dimension. The second Randall-Sundrum(RS2)
scenario\cite{rs99-2} is followed from RS1 by remoting one of the brane 
to infinity. The most remarkable feature of RS2 scenario is that
it leads to a massless graviton localized on the 3-brane at the linearized
fluctuation level\cite{garr00,duff00}. In this paper we will explore
the localized RS graviton problem at RS1 and RS2 from the viewpoint of the 
singular quantum mechanics.

In addition to its good features on hierarchy and localized graviton problem,
RS picture supports a non-static cosmological 
solution\cite{bine99,csa99,cline99} which leads to the conventional 
Friedmann equation if one introduces the bulk and brane cosmological constants
and imposes a particular fine-tuning condition between them. 
Furthermore, RS scenario is also applied to the cosmological constant
hierarchy\cite{kim01,alex01} and black hole physics\cite{cham00,emp00,gidd00}.

The bulk spacetime of RS scenario is two copies of $AdS_5$ glued in a 
$Z_2$-symmetric way along a boundary which is interpreted as the 3-brane
world-volume. It is explicitly seen by examing the line elements;
\begin{eqnarray}
\label{line}
ds^2&=&e^{-2kr_c |\phi|} \eta_{\mu \nu} dx^{\mu} dx^{\nu} + r_c^2
d\phi^2
\hspace{2.0cm} \mbox{(RS1)}     \\   \nonumber
ds^2&=&e^{-2k |y|} \eta_{\mu \nu} dx^{\mu} dx^{\nu} + dy^2
\hspace{2.7cm} \mbox{(RS2)}
\end{eqnarray}
where $|\phi| \leq \pi$ and $|y| < \infty$. The parameter $r_c$ is a radius of 
the compactified fifth dimension. The anology of RS scenario to
$AdS$/CFT\cite{mal98} enables us to explore the finite temperature effect
in RS brane-world scenario by extending $AdS_5$ bulk spacetime to 
Schwarzschild-$AdS_5$\cite{park01-1,park01-2}. 

Inserting the small fluctuation
equations
\begin{eqnarray}
\label{fluctua}
ds^2&=&\left(e^{-2kr_c |\phi|} \eta_{\mu \nu} + h_{\mu \nu}(x, \phi) \right)
dx^{\mu} dx^{\nu} + r_c^2 d\phi^2
\hspace{1.5cm} \mbox{(RS1)}     \\   \nonumber
ds^2&=&\left(e^{-2k |y|} \eta_{\mu \nu} + h_{\mu \nu}(y) \right)
dx^{\mu} dx^{\nu} + dy^2
\hspace{2.7cm} \mbox{(RS2)}
\end{eqnarray}
to $5d$ Einstein equation one can derive a gravitational fluctuation equation
\begin{eqnarray}
\label{grafluc}
& &\hat{H}_{RS} \hat{\psi}(z) = \frac{m^2}{2} \hat{\psi}(z)  \\  \nonumber
& &\hat{H}_{RS} = -\frac{1}{2} \partial_z^2 + V_i(z)
\end{eqnarray}
where $i=1.2$ represents $i^{th}$ RS scenario. For each RS scenario the 
potential becomes
\begin{eqnarray}
\label{potential}
V_1(z)&=&\frac{15 k^2}{8(k|z| + 1)^2} - \frac{3}{2} k
\left[ \delta(z) - \delta(z - z_0) \right]   \\   \nonumber
V_2(z)&=&\frac{15 k^2}{8(k|z| + 1)^2} - \frac{3}{2} k \delta(z)
\end{eqnarray}
where $z_0 = (e^{k r_c \pi} - 1) / k$. The function $\hat{\psi}(z)$ is related 
to the linearized gravitational field $h$ as follows
\begin{eqnarray}
\label{relation1}
h(x, \phi)&=& e^{-\frac{k}{2} r_c |\phi|} \hat{\psi} e^{ipx}
\hspace{2.0cm} \mbox{(RS1)}     \\   \nonumber
h(x, y)&=&e^{-\frac{k}{2} |y|} \hat{\psi} e^{ipx}
\hspace{2.3cm} \mbox{(RS2)}
\end{eqnarray}
where $m^2 = -p^2$ and
\begin{eqnarray}
\label{changeofv}
z&=&\epsilon(\phi) \frac{e^{k r_c |\phi|} - 1}{k}  
\hspace{2.8cm} \mbox{(RS1)}    \\   \nonumber
z&=&\epsilon(y) \frac{e^{k|y|} - 1}{k}.
\hspace{3.1cm} \mbox{(RS2)}
\end{eqnarray}
Since all components are same, Lorentz indicies $\mu$ and $\nu$ are suppressed
in Eq.(\ref{relation1}). 

When deriving the linearized fluctuation equation (\ref{grafluc}), we 
have used
the RS gauge choice
\begin{equation}
\label{gauge}
h^{\mu}_{\nu,\mu} = h^{\mu}_{\mu} = 0, 
\hspace{2.0cm}
h_{55} = h_{\mu 5} = 0
\end{equation}
for each RS scenario. This gauge choice, however, generally generates a
non-trivial bending effect on the brane\cite{garr00}. The bending effect
usually makes the linearized fluctuation equation (\ref{grafluc}) to be 
non-homogeneous form, {\it i.e.} $(\hat{H}_{RS} - m^2/2) \hat{\psi} \neq 0$.
Thus, inclusion of the bending effect makes the stroy to be more
complicated. In this paper we will not consider the bending effect for
simplicity.

The linearized fluctuation equation (\ref{grafluc}) looks like usual
Schr\"{o}dinger equation. From the purely mathematical point of view
the Hamiltonian operator $\hat{H}_{RS}$ in Eq.(\ref{grafluc}) is a singular
operator due to the singular $\delta$-function potential in $V_i$. In the
path-integral framework\cite{fey65,schul81} the $1d$ $\delta$-function 
potential was treated by Schulman about one and half decades ago as 
follows\cite{schul86}.

Let us consider $1d$ Hamiltonian
\begin{equation}
\label{example1}
H = H_V + v \delta(x)
\end{equation}
where 
\begin{equation}
\label{example2}
H_V = \frac{p^2}{2} + V(x).
\end{equation}
It is well-known that the Euclidean propagator $G[x_1, x_2; t]$ for $H$ obeys
the following integral equation
\begin{equation}
\label{inteq}
G[x_1, x_2; t] = G_V[x_1, x_2: t] - v \int_0^t ds \int dx
G_V[x_1, x; t-s] \delta(x) G[x, x_2; s]
\end{equation}
where $G_V[x_1, x_2: t]$ is an Euclidean propagator for $H_V$. The Euclidean
propagator $G[x_1, x_2; t]$ is related to the usual Feynman propagator(or 
Kernel) $K[x_1, x_2; t]$ as follows;
\begin{equation}
\label{kerneldef}
K[x_1, x_2; t] = G[x_1, x_2; it].
\end{equation}
Taking a Laplace transform 
\begin{equation}
\label{lapdef}
\hat{f} \equiv {\cal L} f(t) \equiv \int_0^{\infty} dt e^{-Et} f(t)
\end{equation}
to both sides of Eq.(\ref{inteq}) yields 
\begin{equation}
\label{example3}
\hat{G}[x_1, x_2; E] = \hat{G}_V[x_1, x_2; E] - v \hat{G}_V[x_1, 0; E]
\hat{G}[0, x_2; E]
\end{equation}
which supports a solution
\begin{equation}
\label{example4}
\hat{G}[0, x_2; E] = \frac{\hat{G}_V[0, x_2; E]}{1 + v \hat{G}_V[0, 0, E]}.
\end{equation}
Inserting Eq.(\ref{example4}) into Eq.(\ref{example3}) again completes
Schulman's procedure;
\begin{equation}
\label{schulman}
\hat{G}[x_1, x_2; E] = \hat{G}_V[x_1, x_2; E] - 
\frac{\hat{G}_V[x_1, 0; E] \hat{G}_V[0, x_2; E]}
     {\frac{1}{v} + \hat{G}_V[0, 0; E]}.
\end{equation}
The usual energy-dependent Green's function $\hat{K}[x_1, x_2; E]$ which is 
a Fourier transform of $K[x_1, x_2; t] \theta(t)$, where $\theta(t)$ is a 
step function, is also evaluated from the corresponding fixed-energy
amplitude $\hat{G}[x_1, x_2; E]$ by a relation
\begin{equation}
\label{example5}
\hat{K}[x_1, x_2; E] = -i \hat{G}[x_1, x_2; -E]
\end{equation}
where $-E$ in $\hat{G}$ is a usual Euclidean nature. Of course, one can 
compute the Feynman propagator by taking an inverse Laplace transform 
to $\hat{G}[x_1, x_2; E]$  and using a relation (\ref{kerneldef}).

Extension of Schulman's procedure to higher dimensional cases is not
straightforward due to the infinity arising from origin. In these cases 
we have to modify Eq.(\ref{schulman}) appropriately to escape the 
ultraviolet divergence\cite{park95}. Especially, $2d$ case is very
interesting because a lot of non-trivial effects are involved in $2d$ 
$\delta$-function potential such as scale anomaly and dimensional 
transmutation. In Ref.\cite{jack91} Jackiw explored the $2d$ 
$\delta$-function potential system by making use of the physically-oriented
coupling constant renormalization and the mathematically-oriented
self-adjoint extension\cite{capri85,albev88}. He also derived the relation 
of the
renormalized coupling constant to the self-adjoint extension parameter. His
result is generalized within a path-integral or Green's function 
formalism in Ref.\cite{park95,gros93,gros95,park96,park98}.

The purpose of this paper is to examine the property of the localized 
gravity in RS1 and RS2 scenario by treating Eq.(\ref{grafluc}) as a 
Schr\"{o}dinger equation. In this paper we adopt $AdS$/CFT setting, {\it i.e.}
single copy of $AdS_5$ spacetime with a singular brane on the boundary.
The $AdS$/CFT setting generates non-trivial constraints. For RS1 and RS2
it generates $1d$ box($0\leq \phi \leq \pi$) and half-line($0 \leq y < \infty$)
constraints respectively. These constraints makes the fixed-energy
amplitude for $\hat{H}_{RS}$ to be crucially dependent on the 
boundary conditions(BCs). The combination of these constraints with a singular
$\delta$-function potential makes the situation to be complicated. The 
fascinating fact is that Dirichlet BC requires a coupling constant 
renormalization to lead a non-trivial fixed-energy amplitude although our case 
is one-dimensional singular quantum mechanics. 

The paper is organized as follows. In section 2 we will consider the free
particle case with an half-line constraint and $\delta$-function potential 
at the boundary as a toy model of RS2 case. In this section we will show 
how BCs play important roles in this simple singular quantum mechanics. Also
we will show why coupling constant renormalization is necessary
to lead a non-trivial modification in the fixed-energy amplitude at 
Dirichlet BC. In section 3 we will compute the fixed-energy amplitude
for RS2\cite{park01-3} which depends on a free parameter $\xi$ where
$\xi=0$ and $\xi=1$ correspond respectively to pure Dirichlet and 
pure Neumann BC cases. We will derive in this section the general 
criterion in the parameter space for the localized graviton on the 3-brane to 
be massless. We will also show that the massless graviton at $\xi = 0$ is 
followed via a coupling constant renormalization. In section 4 we will 
consider the free particle case with an $1d$-box constraint and 
$\delta$-function potentials at the both boundaries as a toy model of RS1.
The final expression is dependent on the two free parameters
$\xi_1$ and $\xi_2$ which parametrize the BCs arising at both 
boundaries of $1d$ box.
In section 5 we will compute the fixed-energy amplitude for RS1 which 
depends on two free parameters $\xi_1$ and $\xi_2$. We will show in this
section that there is no localized massless graviton on both branes.
For positive-tension brane, however, the massless graviton can appear when
the width of $1d$ box is infinity, which is essentially identical to RS2. 
For the negative-tension brane our numerical calculation indicates there is 
no localized massless graviton regardless of the size of $1d$ box.
In final 
section a brief conclusion is given.

\section{Toy Model 1: Free Particle on a half-line with $\delta$-function 
Potential} 
In this section as a toy model of RS2 we will examine Green's function
for the free particle system defined on a half-line($x \geq 0$) with 
$\delta$-function potential whose Hamiltonian is 
\begin{equation}
\label{toy1hamil1}
\hat{H} = \hat{H}_0^> - v \delta(x)
\end{equation}
where $\hat{H}_0^>$ is a free particle Hamiltonian with the half-line
constraint, {\it i.e.}
\begin{equation}
\label{toy1hamil2}
\hat{H}_0^> = -\frac{1}{2} \partial_x^2
\hspace{2.0cm}
(x \geq 0).
\end{equation}
Of course the main problem in this model is how to compute the fixed-energy
amplitude for $\hat{H}_0^>$. 
Once this is completed, one can derive a fixed-energy amplitude for 
$\hat{H}$ by employing the Schulman procedure described in the previous
section. 

We start with a fixed-energy amplitude $\hat{G}_F[x, y; E]$ for free particle
without any constraint
\begin{equation}
\label{free1}
\hat{G}_F[x, y; E] = \frac{e^{-\sqrt{2E} |x - y|}}{\sqrt{2E}}.
\end{equation}
Then, the fixed-energy amplitude 
for $\hat{H}_0^>$ can be computed as follows from $\hat{G}_F[x, y; E]$.
First, we have to note that the fixed-energy amplitude for 
$\hat{H}_0^>$ is dependent on BC at $x = 0$ arising due to the half-line
constraint. The usual Dirichlet or Neumann BCs at $x = 0$
are properly incorporated into the path-integral formalism
using $\delta$- and $\delta^{\prime}$-function potentials
with infinite coupling constant\cite{gros93,gros95};
\begin{eqnarray}
\label{deldel}
\hat{G}_F^D[a, b; E]&=&\hat{G}_F[a, b; E] - 
\frac{\hat{G}_F[a, 0; E] \hat{G}_F[0, b; E]}{\hat{G}_F[0^+, 0; E]}
                                              \\   \nonumber
\hat{G}_F^N[a, b; E]&=&\hat{G}_F[a, b; E] -
\frac{\hat{G}_{F, b}[a, 0; E] \hat{G}_{F, a}[0, b; E]}{\hat{G}_{F, ab}[0^+, 0; E]}.
\end{eqnarray}
where the superscripts $D$ and $N$ stand for Dirichlet and Neumann 
respectively. The explicit calculation shows 
\begin{eqnarray}
\label{toy1dn}
\hat{G}_F^D[a, b; E]&=&\frac{e^{-\sqrt{2E} |x - y|}}{\sqrt{2E}} - 
\frac{e^{-\sqrt{2E}(|a| + |b|)}}{\sqrt{2E}}   \\   \nonumber
\hat{G}_F^N[a, b; E]&=&\frac{e^{-\sqrt{2E} |x - y|}}{\sqrt{2E}} + 
\frac{\epsilon(a) \epsilon(b) e^{-\sqrt{2E}(|a| + |b|)}}{\sqrt{2E}}.
\end{eqnarray}
One can show easily $\hat{G}_F^D$ and $\hat{G}_F^N$ satisfy the following BCs;
\begin{eqnarray}
\label{toy1bcs}
\hat{G}_F^D[a, 0; E]&=&\hat{G}_F^D[0, b; E] = 0
                                   \\   \nonumber
\hat{G}_{F, b}^N[a, 0; E]&=&\hat{G}_{F, a}[0, b; E] = 0.
\end{eqnarray}
Then, the general fixed-energy amplitude for $\hat{H}_0^>$ can be obtained
by linearly combining $\hat{G}_F^D$ and $\hat{G}_F^N$;
\begin{equation}
\label{toy1gen1}
\hat{G}_F^{\xi}[a, b; E] = \xi \hat{G}_F^N[a, b; E] + (1 - \xi)
\hat{G}_F^D[a, b; E]
\end{equation}
where $\xi(0 \leq \xi \leq 1)$ is real parameter parametrizing the BCs at the 
origin. Of course $\xi=0$ and $\xi=1$ represent the pure Dirichlet and 
pure Neumann BCs respectively. Another interesting case is $\xi= 1/2$, in
which the contribution of Neumann and Dirichlet have an equal weighting
factors. Since the fixed-energy amplitude $\hat{G}_F^{\xi}$ is in general 
expressed in terms of eigenvalues $E_n$ and eigenfunctions $\phi_n$ of 
$\hat{H}_0^>$ as follows
\begin{equation}
\label{toy1eigen}
\hat{G}_F^{\xi}[a, b; E] = \sum_n \frac{\phi_n(a) \phi_n^{\ast}(b)}{E - E_n},
\end{equation} 
the $\xi=1/2$ case should correspond to the free particle case without any
constraint at the origin. 

Following Schulman procedure one can calculate the fixed-energy
amplitude $\hat{G}^{\xi}$ for $\hat{H}$ from $\hat{G}_F^{\xi}$ as follows;
\begin{equation}
\label{toy1modi}
\Delta \hat{G}^{\xi}[a, b; E] \equiv
\hat{G}^{\xi}[a, b; E] - \hat{G}_F^{\xi}[a, b; E] = 
\frac{4 \xi^2}{\frac{\sqrt{2E}}{v} - 2 \xi}
\frac{e^{-\sqrt{2E}(|a| + |b|)}}{\sqrt{2E}}.
\end{equation}
At $\xi = 1$ and $\xi=1/2$ the fixed-energy amplitudes are 
simply reduced to
\begin{eqnarray}
\label{toy1nrs}
\hat{G}^{\xi=1}[a, b; E]&=& \frac{e^{-\sqrt{2E} |a -b|}}{\sqrt{2E}} +
\frac{\frac{\sqrt{2E}}{v} + 2}{\frac{\sqrt{2E}}{v} - 2}
\frac{e^{-\sqrt{2E} (|a| + |b|)}}{\sqrt{2E}}    \\   \nonumber
\hat{G}^{\xi=\frac{1}{2}}[a, b; E]&=& \frac{e^{-\sqrt{2E} |a -b|}}{\sqrt{2E}} +
\frac{e^{-\sqrt{2E} (|a| + |b|)}}{\sqrt{2E} \left(\frac{\sqrt{2E}}{v} - 1\right)
                                   }
\end{eqnarray}
and the corresponding bound state energies $B(\xi)$ arising due to the 
$\delta$-function potential are
\begin{eqnarray}
\label{toy1bound1}
B(\xi = 1)&=& -2 v^2   \\    \nonumber
B(\xi = \frac{1}{2})&=& = - \frac{v^2}{2}.
\end{eqnarray}

Finally, let us consider $\xi = 0$ case. In this case Eq.(\ref{toy1modi})
shows that the modification term $\Delta \hat{G}^{\xi = 0}$ vanishes. This 
means the $\delta$-function potential in Eq.(\ref{toy1hamil1}) does not
play any important role. In fact this is obvious if we consider the fact
that at $\xi = 0$ the Hamiltonian $\hat{H}^>_0$ describes the free
particle system plus $\lim_{\alpha \rightarrow \infty} \alpha \delta(x)$ which
makes the half-line constraint. Thus, the $\delta$-function potential in
eq.(\ref{toy1hamil1}) is absorbed to $\hat{H}^>_0$.

Even in this case, however, one can derive a non-trivial fixed-energy 
amplitude under the assumption that $v$ is infinite bare coupling constant 
by adopting the coupling constant renormalization. To show this explicitly
we re-express the modification term $\Delta \hat{G}^{\xi = 0}$ as folllows;
\begin{equation}
\label{toy1dmodi}
\Delta \hat{G}^{\xi = 0}[a, b; E] = 
\lim_{\epsilon \rightarrow 0^+}
\frac{\hat{G}_F^{\xi = 0}[a, \epsilon; E] \hat{G}_F^{\xi = 0}[\epsilon, b; E]}
     {\frac{1}{v} - \hat{G}_F^{\xi = 0}[\epsilon, \epsilon; E]}.
\end{equation}
Expanding the denominator and numerator separately one can conclude
\begin{equation}
\label{toy1dmodi2}
\Delta \hat{G}^{\xi = 0}[a, b; E] =
\frac{2}{\sqrt{2E} - v^{ren}} e^{-\sqrt{2E}(|a| + |b|)}
\end{equation}
where the renormalized coupling constant $v^{ren}$ is defined as 
\begin{equation}
\label{toy1rencop}
v^{ren} = \frac{1}{2 \epsilon^2} \left( 2 \epsilon - \frac{1}{v} \right).
\end{equation}
It is easy to show that $v^{ren}$ has a same dimension with the bare
coupling constant $v$. Following the philosophy of renormalization we regard
$v^{ren}$ as a finite quantity. Combining Eq.(\ref{toy1dn}) and 
Eq.(\ref{toy1dmodi2}) we get finally
\begin{equation}
\label{toy1dfinal}
\hat{G}^{\xi = 0}[a, b; E] = \frac{e^{-\sqrt{2E} |a - b|}}{\sqrt{2E}}
+ \frac{\frac{\sqrt{2E}}{v^{ren}} + 1}
       {\frac{\sqrt{2E}}{v^{ren}} - 1}
\frac{e^{-\sqrt{2E}(|a| + |b|)}}{\sqrt{2E}}
\end{equation}
whose bound state energy is $B(\xi=0) = - (v^{ren})^2 / 2$.

In the next section we will apply the analysis in this toy model to the
RS2 scenario.

\section{Fixed-Energy Amplitude for RS2}
Recently, one of the present authors computed the fixed-energy amplitude for 
RS2 at Ref.\cite{park01-3} which will be reviewed in this section briefly. 
Furthermore we will derive the general condition in the parameter space for
the appearance of the localized massless graviton.

The Hamiltonian for RS2 can be read from Eq.(\ref{grafluc}) and (\ref{potential})
easily;
\begin{eqnarray}
\label{rs2hamil1}
\hat{H}_{RS2}&=&\hat{H}_0 - v \delta(z)    \\   \nonumber
\hat{H}_0&=& -\frac{1}{2} \partial_z^2 + 
\frac{g}{(|z| + c)^2}.
\end{eqnarray}
Of course, we can obtain the exact RS2 Hamiltonian by letting $g=15/8$, 
$c = 1/k \equiv R$ and $v = 3k / 2$, where $R$ is the radius of 
$AdS_5$. In this section, however, we do not require them from the 
beginning. In other words we will compute the fixed-energy amplitude 
for arbitrary $g$, $c$, and $v$ when the half-line constraint ($z \geq 0$)
is imposed. This will give us the general condition for the localized graviton
on the brane to be massless.

The half-line constraint makes $\hat{H}_0$ in Eq.(\ref{rs2hamil1}) to be a 
following simple form;
\begin{equation}
\label{rs2hamil2}
\hat{H}_0 = -\frac{1}{2} \partial_x^2 + \frac{g}{x^2}
\end{equation}
where $x = z + c$. Thus our half-line constraint $z \geq 0$ is changed into
$x \geq c$. If $c = 0$, the Euclidean propagator $G_{>0}[a, b; t]$ and the 
corresponding fixed-energy amplitude $\hat{G}_{>0}[a, b; E]$ for Hamiltonian
(\ref{rs2hamil2}) are given at Ref.\cite{schul81};
\begin{eqnarray}
\label{rs2schul}
G_{>0}[a, b; t]&=&\frac{\sqrt{a b}}{t} e^{-\frac{a^2 + b^2}{2 t}} 
I_{\gamma}\left( \frac{a b}{t} \right)    \\   \nonumber
\hat{G}_{>0}[a, b; E]&=& 2 \sqrt{a b} 
I_{\gamma}\left( \sqrt{\frac{E}{2}} [(a + b) - |a - b|] \right)
K_{\gamma}\left( \sqrt{\frac{E}{2}} [(a + b) + |a - b|] \right)
\end{eqnarray}
where $I_{\gamma}(z)$ and $K_{\gamma}(z)$ are the usual modified
Bessel functions, and $\gamma = \sqrt{1 + 8 g} / 2$.

The main problem for the computation of the fixed energy amplitude
for $\hat{H}_0$ in Eq.(\ref{rs2hamil2}) is how to adopt an asymmetric
constraint $x \geq c$ in terms of $x$. However, this is already explained
at the previous section by introducing an infinite energy barrier.
In the asymmetric barrier the fixed-energy amplitude will be dependent
on the BC at $x = c$. Hence, the final form will be one parameter 
family type
\begin{equation}
\label{rs2onepara}
\hat{G}_0^{\xi}[a, b; E] = \xi \hat{G}_0^N[a, b; E] + 
(1 - \xi) \hat{G}_0^D[a, b; E]
\end{equation}
where $\hat{G}_0^N$ and $\hat{G}_0^D$ are the fixed-energy amplitudes which 
obey the Neumann and Dirichlet BCs respectively at $x = c$. Of course, 
$\hat{G}_0^N$ and $\hat{G}_0^D$ can be calculated from 
$\hat{G}_{>0}[a, b; E]$ introducing $\delta$- and $\delta^{\prime}$-functions 
as we did in the previous section;
\begin{eqnarray}
\label{rs2gdgn}
\hat{G}_0^D[a, b; E]&=&\hat{G}_{>0}[a, b; E] - 
\frac{\hat{G}_{>0}[a, c; E] \hat{G}_{>0}[c, b; E]}
     {\hat{G}_{>0}[c^+, c; E]}          \\    \nonumber
\hat{G}_0^N[a, b; E]&=&\hat{G}_{>0}[a, b; E] -
\frac{\hat{G}_{>0, b}[a, c; E] \hat{G}_{>0, a}[c, b; E]}
     {\hat{G}_{>0, ab}[c^+, c; E]}.
\end{eqnarray}

Inserting Eq.(\ref{rs2schul}) into Eq.(\ref{rs2gdgn}) one can derive the
explicit forms of $\hat{G}_0^D$ and $\hat{G}_0^N$;
\begin{eqnarray}
\label{rs2gdgn2}
\hat{G}_0^D[a, b; E]&=&\hat{G}_{>0}[a, b; E] - 2 \sqrt{a b}
\frac{I_{\gamma}(\sqrt{2E} c)}{K_{\gamma}(\sqrt{2E} c)}
K_{\gamma}(\sqrt{2E} a) K_{\gamma}(\sqrt{2E} b)
                                               \\   \nonumber
\hat{G}_0^N[a, b; E]&=&\hat{G}_{>0}[a, b; E] + 2 \sqrt{a b}
\frac{f_I(c, E)}{f_K(c, E)} 
K_{\gamma}(\sqrt{2E} a) K_{\gamma}(\sqrt{2E} b)
\end{eqnarray}
where
\begin{eqnarray}
\label{fkfi}
f_K(x, E)&=& \frac{\gamma - \frac{1}{2}}{\sqrt{2E} x} K_{\gamma}(\sqrt{2E} x)
+K_{\gamma - 1}(\sqrt{2E} x)    \\   \nonumber
f_I(x, E)&=&I_{\gamma - 1}(\sqrt{2E} x)
- \frac{\gamma - \frac{1}{2}}{\sqrt{2E} x} I_{\gamma}(\sqrt{2E} x).
\end{eqnarray}
The useful relation which will be used frequently is 
\begin{equation}
\label{frequent}
f_K(x, E) I_{\gamma}(\sqrt{2E} x) + f_I(x, E) K_{\gamma}(\sqrt{2E} x)
= \frac{1}{\sqrt{2E} x}.
\end{equation}

Following Schulman procedure it is straightforward to derive a fixed-energy
amplitude for $\hat{H}_{RS2}$;
\begin{equation}
\label{rs2schulman}
\hat{G}_{RS2}[a, b; E] = \hat{G}_0^{\xi}[a, b; E]
+ \frac{\hat{G}_0^{\xi}[a, c; E]\hat{G}_0^{\xi}[c, b; E]}
       {\frac{1}{v} - \hat{G}_0^{\xi}[c^+, c; E]}
\end{equation}
where $\hat{G}_0^{\xi}$ is given in Eq.(\ref{rs2onepara}). The most convenient
form of $\hat{G}_{RS2}$ is  
\begin{eqnarray}
\label{rs2conven}
\hat{G}_{RS2}[a, b; E]&=& \hat{G}_0^D[a, b; E]
+ \frac{\sqrt{a b}}{c v}
\frac{K_{\gamma}(\sqrt{2E} a) K_{\gamma}(\sqrt{2E} b)}
     {K_{\gamma}^2(\sqrt{2E} c)}   \\  \nonumber
& &
\hspace{3.0cm} \times
\left[ \left( \frac{\gamma - \frac{1}{2}}{2\xi c v} - 1 \right)
       + \frac{\sqrt{2E}}{2 \xi v}
        \frac{K_{\gamma - 1}(\sqrt{2E} c)}
             {K_{\gamma}(\sqrt{2E} c)}         \right]^{-1}.
\end{eqnarray}
Since $\hat{G}_0^D$ satisfies the usual Dirichlet BC, {\it i.e.}
$\hat{G}_0^D[a, c; E] = \hat{G}_0^D[c, b; E] = 0$, the fixed-energy
amplitude on the brane is simply reduced to
\begin{equation}
\label{rs2onbrane}
\hat{G}_{RS2}[c, c; E] =
\left[ \left( \frac{\gamma - \frac{1}{2}}{2 \xi c} - v \right)
       + \frac{\sqrt{2E}}{2 \xi}
        \frac{K_{\gamma - 1}(\sqrt{2E} c)}
             {K_{\gamma}(\sqrt{2E} c)}         \right]^{-1}.
\end{equation}
If $(\gamma - 1/2) / (2 \xi c) - v = 0$, $\hat{G}_{RS2}[c, c; E]$
becomes
\begin{equation}
\label{rs2simple}
\hat{G}_{RS2}[c, c; E] = \Delta_0 + \Delta_{KK}
\end{equation}
where
\begin{eqnarray}
\label{d0dkk}
\Delta_0&=&\frac{2 \xi (\gamma - 1)}{c E}    \\   \nonumber
\Delta_{KK}&=&\frac{2 \xi}{\sqrt{2E}}
              \frac{K_{\gamma - 2}(\sqrt{2E} c)}
                   {K_{\gamma - 1}(\sqrt{2E} c)}.
\end{eqnarray}
Of course, $\Delta_0$ and $\Delta_{KK}$ represent the zero mode and the higher
Kaluza-Klein excitations respectively. This means that the condition
for the localized graviton to be massless is 
\begin{equation}
\label{rs2condi}
\frac{\gamma - \frac{1}{2}}{2 \xi c} - v = 0.
\end{equation}
At the RS limit $\gamma = 2$, $c = R$ and $v = 3 / 2 R$ this condition 
really holds at $\xi = 1/2$. But for other valus of $\xi$ except
$\xi = 0$ one can also obtain the massless graviton by changing $c$,
$g$, and $v$ appropriately to obey Eq.(\ref{rs2condi}).

At the pure Dirichlet BC case($\xi = 0$) Eq.(\ref{rs2condi}) cannot hold 
unless $v = \infty$. Thus in this case we can get the massless graviton
via a coupling constant renormalization as we did in the previous 
section. 

To show explicitly we have to re-write Eq.(\ref{rs2schulman}) by introducing 
a positive infinitesimal parameter $\epsilon$;
\begin{equation}
\label{rs2d1}
\hat{G}_{RS2}^{\xi=0}[a, b; E] = \hat{G}_0^D[a, b; E]
+ \lim_{\epsilon \rightarrow 0^+}
\frac{\hat{G}_0^D[a, c + \epsilon; E] \hat{G}_0^D[c + \epsilon, b; E]}
     {\frac{1}{v} - \hat{G}_0^D[c + \epsilon, c + \epsilon; E]}.
\end{equation}
Explicit calculation shows that even in this case one can derive a massless
graviton when $v^{ren} = -3 / (2 R)$, where $v^{ren}$ is a renormalized
coupling constant defined as 
\begin{equation}
\label{rs2vren}
v^{ren} = \frac{1}{2 \epsilon^2}
\left( \frac{1}{v} - 2 \epsilon \right).
\end{equation}

The coupling constant renormalization procedure in RS2 is in detail
explained in Ref.\cite{park01-3}. 
In this case the fixed-energy amplitude and the corresponding gravitational
potential is exactly same with that of the original RS result
when $v^{ren} = -3 / (2 R)$. This result may provide us the compromise
of the massless graviton with a small cosmological constant\cite{park01-3}.

\section{Toy Model 2: Free Particle in a Box with $\delta$-function
          Potentials}
In this section as a toy model of RS1 we will examine Green's function
for the free particle system in a $1d$ box($0 \leq x \leq L$) with 
$\delta$-function potentials at both end points.

The Hamiltonian for this system is 
\begin{equation}
\label{toy2hamil1}
\hat{H}_{\delta}^{Box} = \hat{H}_0^{Box} - v_1 \delta(x) + v_2 \delta(x - L)
\end{equation}
where
\begin{equation}
\label{toy2hamil2}
\hat{H}_0^{Box} = - \frac{1}{2} \partial_x^2
\hspace{2.0cm}
(0 \leq x \leq L).
\end{equation}
The main problem in this toy model is of course to derive a fixed-energy
amplitude $\hat{G}_0^{Box}$ for $\hat{H}_0^{Box}$. Once $\hat{G}_0^{Box}$
is obtained, the fixed-energy amplitude $\hat{G}_{\delta}^{Box}$ for the 
total Hamiltonian $\hat{H}_{\delta}^{Box}$ is straightforwardly obtained
by performing the Schulman procedure twice;
\begin{equation}
\label{toy2schul1}
\hat{G}_{\delta}^{Box}[a, b; E] =
\hat{{\cal G}}_0^{Box}[a, b; E] - 
\frac{\hat{{\cal G}}_0^{Box}[a, L; E] \hat{{\cal G}}_0^{Box}[L, b; E]}
     {\frac{1}{v_2} + \hat{{\cal G}}_0^{Box}[L, L^-; E]} 
\end{equation}
where
\begin{equation}
\label{toy2schul2}
\hat{{\cal G}}_0^{Box}[a, b; E] = \hat{G}_0^{Box}[a, b; E]
+ \frac{\hat{G}_0^{Box}[a, 0; E] \hat{G}_0^{Box}[0, b; E]}
       {\frac{1}{v_1} - \hat{G}_0^{Box}[0^+, 0; E]}.
\end{equation}

The fixed-energy amplitude $\hat{G}_0^{Box}$ for $\hat{H}_0^{Box}$ is also
obtained directly from that for free particle on half-line, {\it i.e.}
$\hat{G}_F^{\xi_1}[a,b; E]$ in Eq.(\ref{toy1gen1}). Of course, the parameter
$\xi_1$ represents the type of BC at $x = 0$. Then, the fixed-energy amplitude
$\hat{G}_0^{Box}$ can be computed by introducing an infinite barrier at
$x = L$ to the half-line constraint system. As we commented in section 2 the
infinite barrier is introduced by $\delta$- and $\delta^{\prime}$-functions
potential at $x = L$ with assumption that the coupling constant is
infinite. Thus the final form of $\hat{G}_0^{Box}$ is dependent on the two
parameters as follows;
\begin{equation}
\label{toy2box1}
\hat{G}_0^{Box}[a, b; E] = \xi_2 \hat{G}_0^{\xi_1, N}[a, b; E]
+ (1 - \xi_2) \hat{G}_0^{\xi_1, D}[a, b; E]
\end{equation}
where
\begin{eqnarray}
\label{toy2gdgn}
\hat{G}_0^{\xi_1, D}[a, b; E]&=&\hat{G}_F^{\xi_1}[a, b; E] - 
\frac{\hat{G}_F^{\xi_1}[a, L; E] \hat{G}_F^{\xi_1}[L, b; E]}
     {\hat{G}_F^{\xi_1}[L, L^-; E]}
                                        \\   \nonumber
\hat{G}_0^{\xi_1, N}[a, b; E]&=&\hat{G}_F^{\xi_1}[a, b; E] -
\frac{\hat{G}_{F,b}^{\xi_1}[a, L; E] \hat{G}_{F, a}^{\xi_1}[L, b; E]}
     {\hat{G}_{F, ab}^{\xi_1}[L, L^-; E]}.
\end{eqnarray}
Of course the parameter $\xi_2$ in Eq.(\ref{toy2box1}) parametrizes the
various BCs at $x = L$. 

Explicit calculation shows 
\begin{eqnarray}
\label{toy2explicit}
\hat{G}_0^{\xi_1, D}[a, b; E]&=&
\frac{1}
     {\sqrt{2E} [\xi_1 \cosh \sqrt{2E} L + (1 - \xi_1) \sinh \sqrt{2E} L]}
                                                         \\   \nonumber
& & \times
\Bigg[ \xi_1 \bigg\{\sinh \sqrt{2E} (L - |a - b|) - \sinh \sqrt{2E} ((a + b) 
                                                             - L) \bigg\}
                                                   \\  \nonumber
& & \hspace{1.0cm}
      + (1 - \xi_1) \bigg\{\cosh \sqrt{2E} (L - |a - b|) - \cosh 
                                         \sqrt{2E} ((a + b) - L) \bigg\}
                                                  \Bigg]
                                                     \\  \nonumber
\hat{G}_0^{\xi_1, N}[a, b; E]&=&
\frac{1}
     {\sqrt{2E} [(1- \xi_1)\cosh \sqrt{2E} L + xi_1 \sinh \sqrt{2E} L]}
                                                          \\   \nonumber
& & \times
\Bigg[ \xi_1 \bigg\{\cosh \sqrt{2E} (L - |a - b|) + \cosh \sqrt{2E} ((a + b)
                                                              - L) \bigg\}
                                                    \\  \nonumber
& & \hspace{1.0cm}
+ (1 - \xi_1) \bigg\{\sinh \sqrt{2E} (L - |a - b|) + \sinh 
                                            \sqrt{2E} ((a + b) - L) \bigg\}
                                                    \Bigg].
\end{eqnarray}
Inserting eq.(\ref{toy2explicit}) into Eq.(\ref{toy2box1}) we get
\begin{eqnarray}
\label{toy2box2}
\hat{G}_0^{Box}[a, b; E]&=& \frac{1}{\sqrt{2E}}
\Bigg[ \left( \mu(\xi_1, \xi_2) + \mu(1 - \xi_1, 1 - \xi_2) \right)
                               \cosh \sqrt{2E} (L - |a - b|)
                                              \\   \nonumber
& & \hspace{1.0cm} 
+ \left( \mu(\xi_1, \xi_2) - \mu(1 - \xi_1, 1 - \xi_2) \right)
                               \cosh \sqrt{2E} ((a + b) - L)
                                               \\   \nonumber
& & \hspace{1.0cm} 
+ \left( \nu(\xi_1, \xi_2) + \nu(1 - \xi_1, 1 - \xi_2) \right)
                               \sinh \sqrt{2E} (L - |a - b|)
                                               \\   \nonumber
& & \hspace{1.0cm} 
+ \left( \nu(\xi_1, \xi_2) - \nu(1 - \xi_1, 1 - \xi_2) \right)
                               \sinh \sqrt{2E} ((a + b) - L)
                                                           \Bigg]
\end{eqnarray}                                                         
where
\begin{eqnarray}
\label{toy2munu}
\mu(z, w)&=& \frac{z w}{(1 - z) \cosh \sqrt{2E} L + z \sinh \sqrt{2E} L}
                                                   \\   \nonumber
\nu(z, w)&=& \frac{(1 - z) w}{(1 - z) \cosh \sqrt{2E} L + z \sinh \sqrt{2E} L}.
\end{eqnarray}
It is interesting to note the following special cases;
\begin{eqnarray}
\label{toy2special}
\hat{G}_0^{DD}[a, b; E]&=&
\frac{\cosh \sqrt{2E} (L - |a - b|) - \cosh \sqrt{2E} ((a + b) - L)}
     {\sqrt{2E} \sinh \sqrt{2E} L}
                                           \\   \nonumber
\hat{G}_0^{NN}[a, b; E]&=&
\frac{\cosh \sqrt{2E} (L - |a - b|) + \cosh \sqrt{2E} ((a + b) - L)}
     {\sqrt{2E} \sinh \sqrt{2E} L}
\end{eqnarray}
where the superscript $DD$(or $NN$) stands for 
Dirichlet-Dirichlet(or Neumann-Neumann) BCs at $x = 0$ and $x = L$. 
Similar results to Eq.(\ref{toy2special}) are found at Ref.\cite{gros98}.
Bound state energy spectrum is obtained from poles of $\hat{G}_0^{DD}$ 
and $\hat{G}_0^{NN}$ which indicates $B_n^{DD} = B_n^{NN} = n^2 \pi^2 / 2 L^2$
where $n$ is integer.
Another interesting case is $\xi_1 = \xi_2 = 1/2$ case where $\hat{G}_0^{Box}$
is simply reduced to the free particle case without any constraint,
{\it i.e.} $e^{-\sqrt{2E} |a - b|} / \sqrt{2E}$.

Inserting Eq.(\ref{toy2box2}) into Eq.(\ref{toy2schul2}) and 
subsequently Eq.(\ref{toy2schul1}) we get the final form of 
$\hat{G}_{\delta}^{Box}$ for the Hamiltonian $\hat{H}_{\delta}^{Box}$.
Since final expression is too long, we do not describe it explicitly
in this paper. Instead we will consider two special cases.

The first case we will consider is $\xi_1 = \xi_2 = 1/2$.
In this case Eq.(\ref{toy2schul1}) and Eq.(\ref{toy2schul2}) yield
\begin{eqnarray}
\label{toy2xi1/2}
\hat{G}_{\delta, \xi_1=\xi_2=1/2}^{Box}[a, b; E]&=&
\frac{e^{-\sqrt{2E}|a - b|}}{\sqrt{2E}} + 
\left( \frac{\sqrt{2E}}{v_1} - 1 \right)^{-1}
\frac{e^{-\sqrt{2E}(a + b)}}{\sqrt{2E}}
                                         \\    \nonumber
& & -
\left[ \left( \frac{\sqrt{2E}}{v_1} - 1 \right)
       \left( \frac{\sqrt{2E}}{v_2} + 1 \right)
       + e^{-2 \sqrt{2E} L}                     \right]^{-1}
                                                       \\  \nonumber
& & \times
\Bigg[ \left( \frac{\sqrt{2E}}{v_1} - 1 \right)
       \frac{e^{\sqrt{2E}(a + b)}}{\sqrt{2E}} +
       \frac{e^{\sqrt{2E}(a - b)}}{\sqrt{2E}} +
       \frac{e^{-\sqrt{2E}(a - b)}}{\sqrt{2E}} +
                                                       \\   \nonumber
& & \hspace{6.0cm}
       \left( \frac{\sqrt{2E}}{v_1} - 1 \right)^{-1}
       \frac{e^{-\sqrt{2E}(a + b)}}{\sqrt{2E}}
                                                \Bigg].
\end{eqnarray}

The second case we will consider is $\xi_1 = \xi_2 = 0$. In this case 
$\hat{G}_0^{Box}$ is $\hat{G}_0^{DD}$ in Eq.(\ref{toy2special}). As expected
$\hat{G}_0^{DD}$ satisfies the usual Dirichlet-Dirichlet BCs;
\begin{equation}
\label{toy2ddbcs}
\hat{G}_0^{DD}[0, b; E] = \hat{G}_0^{DD}[a, 0; E]
= \hat{G}_0^{DD}[L, b; E] = \hat{G}_0^{DD}[a, L; E] = 0.
\end{equation}
If, therefore, $v_1$ and $v_2$ are finite, we arrive at a conclusion
$\hat{G}_{\delta}^{Box}[a, b; E] = \hat{G}_0^{Box}[a, b; E]$.

If however, $v_1$ and $v_2$ are infinite and unphysical bare quantities,
one can arrive at different conclusion via the coupling constant
renormalization as we have seen in section 2 and 3. To adopt the 
coupling constant renormalization we introduce the infinitesimal positive 
constant $\epsilon$ as follows;
\begin{eqnarray}
\label{toy2ddepsi}
\hat{G}_0^{Box}[0, 0; E]&\rightarrow&\hat{G}_0^{Box}[\epsilon^-, \epsilon, E]
= 2 \epsilon - 2 \sqrt{2E} \coth \sqrt{2E} L \epsilon^2 + 
{\cal O} (\epsilon^3)
                                         \\   \nonumber
\hat{G}_0^{Box}[L, L; E]&\rightarrow&\hat{G}_0^{Box}[L-\epsilon^-, L-\epsilon, E]
= 2 \epsilon - 2 \sqrt{2E} \coth \sqrt{2E} L \epsilon^2 + 
{\cal O} (\epsilon^3)
                                         \\   \nonumber
\hat{G}_0^{Box}[0, L; E]&\rightarrow&\hat{G}_0^{Box}[\epsilon, L-\epsilon, E] =
\frac{2 \sqrt{2E}}{\sinh \sqrt{2E} L} \epsilon^2 + {\cal O} (\epsilon^3)
                                          \\   \nonumber
\hat{G}_0^{Box}[L, 0; E]&\rightarrow&\hat{G}_0^{Box}[L-\epsilon, \epsilon, E] =
\frac{2 \sqrt{2E}}{\sinh \sqrt{2E} L} \epsilon^2 + {\cal O} (\epsilon^3)
                                          \\   \nonumber
\hat{G}_0^{Box}[a, 0; E]&\rightarrow&\hat{G}_0^{Box}[a, \epsilon, E] =
\frac{2 \sinh \sqrt{2E} (L-a)}{\sinh \sqrt{2E} L} \epsilon + {\cal O} (\epsilon^3)
                                          \\   \nonumber
\hat{G}_0^{Box}[a, L; E]&\rightarrow&\hat{G}_0^{Box}[a, L-\epsilon, E] =
\frac{2 \sinh \sqrt{2E} a}{\sinh \sqrt{2E} L} \epsilon + {\cal O} (\epsilon^3)
                                          \\   \nonumber
\hat{G}_0^{Box}[0, b: E]&\rightarrow&\hat{G}_0^{Box}[\epsilon, b; E] =
\frac{2 \sinh \sqrt{2E} (L-b)}{\sinh \sqrt{2E} L} \epsilon + {\cal O} (\epsilon^3)
                                          \\   \nonumber
\hat{G}_0^{Box}[L, b; E]&\rightarrow&\hat{G}_0^{Box}[L-\epsilon, b; E] =
\frac{2 \sinh \sqrt{2E} b}{\sinh \sqrt{2E} L} \epsilon + {\cal O} (\epsilon^3).
\end{eqnarray}
Inserting Eq.(\ref{toy2ddepsi}) into Eq.(\ref{toy2schul2}) and Eq.(\ref{toy2schul1}), and 
defining the renormalized constants
\begin{eqnarray}
\label{toy2rencoup}
v_1^{ren}&=&\frac{1}{2\epsilon^2}
\left( \frac{1}{v_1} - 2\epsilon \right)    \\   \nonumber
v_2^{ren}&=&\frac{1}{2\epsilon^2}
\left( \frac{1}{v_2} + 2\epsilon \right),
\end{eqnarray}
one can arrive at the following long expression after tedius calculation;
\begin{eqnarray}
\label{toy2final}
\hat{G}_{\delta, \xi_1=\xi_2=0}^{Box}[a, b; E]&=&\hat{G}_0^{DD}[a, b; E] 
                                                    \\   \nonumber
& &
+ 2 (v_1^{ren} + \sqrt{2E} \coth \sqrt{2E}L)^{-1}
\frac{\sinh \sqrt{2E} (L - a) \sinh \sqrt{2E} (L - b)}{\sinh^2 \sqrt{2E} L}
                                                     \\    \nonumber
&-&
2 \left[ (v_1^{ren} + \sqrt{2E} \coth \sqrt{2E} L) (v_2^{ren} - 
                                                \sqrt{2E} \coth \sqrt{2E} L)
       + \frac{2E}{\sinh^2 \sqrt{2E} L}       \right]^{-1}
                                                      \\    \nonumber
&\times&
\Bigg[(v_1^{ren} + \sqrt{2E} \coth \sqrt{2E} L)
      \frac{\sinh \sqrt{2E} a \sinh \sqrt{2E} b}{\sinh^2 \sqrt{2E} L}
                                                       \\   \nonumber
& &   + \sqrt{2E}
      \frac{\sinh \sqrt{2E} a \sinh \sqrt{2E} (L - b) + 
            \sinh \sqrt{2E} (L - a) \sinh \sqrt{2E} b}
           {\sinh^3 \sqrt{2E} L}
                                                        \\   \nonumber
& &   + 2E (v_1^{ren} + \sqrt{2E} \coth \sqrt{2E} L)^{-1}
       \frac{\sinh \sqrt{2E} (L - a) \sinh \sqrt{2E} (L - b)}
            {\sinh^4 \sqrt{2E} L} \Bigg].    
\end{eqnarray}
In the next section we will apply the analysis in this toy model to the
RS1 scenario.

\section{Fixed-energy Amplitude for RS1}
In this section we will examine the fixed-energy amplitude for RS1 whose linear
gravitational fluctuation is given in Eq.(\ref{grafluc}) and Eq.(\ref{potential}).
The Hamiltonian for RS1 can be read from these equations easily
\begin{eqnarray}
\label{rs1hamil1}
\hat{H}_{RS1}&=&\hat{H}_0 - v_1 \delta(z) + v_2 \delta(z - z_0)
                                                   \\   \nonumber
\hat{H}_0&=&-\frac{1}{2} \partial_z^2 + 
\frac{g}{(|z| + c)^2}.
\end{eqnarray}
Of course, the exact RS1 Hamiltonian can be obtained by letting
$g=15/8$, $c = 1/k \equiv R$, $v_1 = v_2 = 3 k /2$ and 
$z_0 = (e^{k r_c \pi} - 1) / k$. As we did in section 3, however, we will
try to examine the fixed energy amplitude for arbitrary parameter as much
as possible.

Next we impose $z$ is non-negative. This means we use the single copy of
$AdS_5$ as a bulk spacetime. In this sense we have a same setting with that of 
$AdS$/CFT. In this setting Hamiltonian $\hat{H}_0$ in Eq.(\ref{rs1hamil1})
becomes
\begin{equation}
\label{rs1hamil2}
\hat{H}_0 = - \frac{1}{2} \partial_x^2 + \frac{g}{x^2}
\hspace{2.0cm}
(c \leq x \leq L)
\end{equation}
where $L = c + z_0$. 

Of course, the main problem is to compute the fixed-energy amplitude
$\hat{G}_0[a, b; E]$ for $\hat{H}_0$ in Eq.(\ref{rs1hamil2}). From 
$\hat{G}_0[a, b; E]$ it is simple to derive the fixed energy amplitude
for $\hat{H}_{RS1}$ by applying the Schulman procedure twice;
\begin{equation}
\label{rs1schul1}
\hat{G}_{RS1}[a, b; E] =
\hat{{\cal G}}_0[a, b; E] - 
\frac{\hat{{\cal G}}_0[a, L; E] \hat{{\cal G}}_0[L, b; E]}
     {\frac{1}{v_2} + \hat{{\cal G}}_0[L, L^-; E]}
\end{equation}
where
\begin{equation}
\label{rs1schul2}
\hat{{\cal G}}_0[a, b; E] =
\hat{G}_0[a, b;E] + 
\frac{\hat{G}_0[a, c; E] \hat{G}_0[c, b; E]}
     {\frac{1}{v_1} - \hat{G}_0[c^+, c; E]}.
\end{equation}

The fixed-energy amplitude for $\hat{H}_0$ is also straightforwardly
obtained from $\hat{G}_0^{\xi_1}$ in Eq.(\ref{rs2onepara}) by introducing 
an infinite barrier at $x = L$ again. Then, the amplitude is dependent on
the two parameters $\xi_1$ and $\xi_2$ which represent the various BCs
at $x = 0$ and $x = L$ respectively;
\begin{eqnarray}
\label{rs1twopara}
\hat{G}_0[a, b; E]&\equiv& \hat{G}_0^{\xi_1, \xi_2}[a, b; E]
                                                \\    \nonumber
&=&
\xi_2 \hat{G}_0^{\xi_1, N}[a, b; E] + (1 - \xi_2)
\hat{G}_0^{\xi_1, D}[a, b; E]
\end{eqnarray}
where
\begin{eqnarray}
\label{rs1gdgn}
\hat{G}_0^{\xi_1, D}[a, b; E]&=&\hat{G}_0^{\xi_1}[a, b; E] -
\frac{\hat{G}_0^{\xi_1}[a, L; E] \hat{G}_0^{\xi_1}[L, b; E]}
     {\hat{G}_0^{\xi_1}[L, L^-; E]}
                                       \\   \nonumber
\hat{G}_0^{\xi_1, N}[a, b; E]&=&\hat{G}_0^{\xi_1}[a, b; E] -
\frac{\hat{G}_{0, b}^{\xi_1}[a, L; E] \hat{G}_{0, a}^{\xi_1}[L, b; E]}
     {\hat{G}_{0, ab}^{\xi_1}[L, L^-; E]}.
\end{eqnarray}
Explicit calculation shows
\begin{eqnarray}
\label{rs1explicit}
\hat{G}_0^{\xi_1}[a, b; E]&=&2\sqrt{ab}
\Bigg[I_{\gamma}(\sqrt{2E} min(a, b)) K_{\gamma}(\sqrt{2E} max(a, b))
                                                       \\   \nonumber
& & \hspace{3.0cm} 
+ g_1(\xi_1,E) K_{\gamma}(\sqrt{2E} a) K_{\gamma}(\sqrt{2E} b)
                                                       \Bigg]
                                                 \\   \nonumber
\hat{G}_0^{\xi_1, D}[a, b; E]&=&\hat{G}_0^{\xi_1}[a, b; E] -
2\sqrt{ab} g_D(\xi_1, E) [I_{\gamma}(\sqrt{2E} a) + 
                          g_1(\xi_1,E) K_{\gamma}(\sqrt{2E} a) ]
                                                  \\   \nonumber
& & \hspace{3.0cm} \times
                         [I_{\gamma}(\sqrt{2E} b) +
                          g_1(\xi_1,E) K_{\gamma}(\sqrt{2E} b)]
                                                 \\    \nonumber
\hat{G}_0^{\xi_1, N}[a, b; E]&=&\hat{G}_0^{\xi_1}[a, b; E] -
2\sqrt{ab} g_N(\xi_1, E) [I_{\gamma}(\sqrt{2E} a) + 
                          g_1(\xi_1,E) K_{\gamma}(\sqrt{2E} a) ]
                                                 \\    \nonumber
& & \hspace{3.0cm} \times
                         [I_{\gamma}(\sqrt{2E} b) +
                          g_1(\xi_1,E) K_{\gamma}(\sqrt{2E} b)]
\end{eqnarray}
where
\begin{eqnarray}
\label{rs1coeff}
g_1(\xi_1,E)&=&\xi_1 \frac{f_I(c,E)}{f_K(c, E)} - 
(1 - \xi_1) \frac{I_{\gamma}(\sqrt{2E} c)}
               {K_{\gamma}(\sqrt{2E} c)}
                                            \\   \nonumber
g_D(\xi_1, E)&=&
\left( g_1(\xi_1,E) + \frac{I_{\gamma}(\sqrt{2E} L)}{K_{\gamma}(\sqrt{2E} L)}
                                                      \right)^{-1}
                                             \\   \nonumber
g_N(\xi_1, E)&=&
\left( g_1(\xi_1,E) - \frac{f_I(L, E)}{f_K(L, E)} \right)^{-1}
\end{eqnarray}
and, $f_K$ and $f_I$ are defined at Eq.(\ref{fkfi}).

Inserting Eq.(\ref{rs1explicit}) into Eq.(\ref{rs1twopara}) one can obtain
the fixed-energy amplitude for Hamiltonian (\ref{rs1hamil2});
\begin{eqnarray}
\label{rs1xi1xi2}
\hat{G}_0^{\xi_1, \xi_2}[a, b; E]&=&\hat{G}_0^{\xi_1}[a, b; E]
- 2 \sqrt{a b} g_{Box}(\xi_1, \xi_2, E)
[I_{\gamma}(\sqrt{2E} a) +
                          g_1(\xi_1,E) K_{\gamma}(\sqrt{2E} a) ]
                                                 \\  \nonumber
& & \hspace{3.0cm} \times
[I_{\gamma}(\sqrt{2E} b) +
                          g_1(\xi_1,E) K_{\gamma}(\sqrt{2E} b)]
\end{eqnarray}
where
\begin{equation}
\label{rs1gbox}
g_{Box}(\xi_1, \xi_2, E) = \xi_2 g_N(\xi_1, E) + (1 - \xi_2)
g_D(\xi_1, E).
\end{equation}
Thus inserting Eq.(\ref{rs1xi1xi2}) into Eq.(\ref{rs1schul2}) and 
subsequently Eq.(\ref{rs1schul1}) we can derive the fixed-energy 
amplitude $\hat{G}_{RS1}[a,b; E]$. The expression is too long
to describe it here. So, we rely on the numerical computation to
check the occurrence of the localized massless graviton.

First, we will check the possibility for the appearance of the 
massless graviton at the brane located in $x = c$. Fig. 1 shows
$m^2 \hat{G}_0^{\xi_1, \xi_2}[c, c; m^2 / 2]$ when $\xi_1 = \xi_2
= 1/2$ and $R = 1$. Of course we have taken RS limit, {\it i.e.}
$c= 1$, $\gamma=2$, and $v_1 = v_2 = 1.5$. In order for the massless
graviton to appear on the brane we need a pole in
$\hat{G}_0^{\xi_1, \xi_2}[c, c; m^2 / 2]$ at $m^2 = 0$. This means the 
numerical value of $m^2 \hat{G}_0^{\xi_1, \xi_2}[c, c; m^2 / 2]$ should be
non-zero and finite at $m^2 \rightarrow 0$. Fig. 1 indicates that 
the zero mass graviton appears only when L is infinitely large. In fact, 
this limit is effectively RS2 scenario. 

Numerical calculation shows that there is no massless graviton on the 
brane located at $x = L$ regardless of $L$ if one chooses 
$\xi_1 = \xi_2 = 1/2$. One may conjecture that the condition (\ref{rs2condi})
for the appearance of the massless graviton in RS2 may be modified
to
\begin{equation}
\label{rs1wrongcondi}
\frac{\gamma - \frac{1}{2}}{2 \xi_2 L} - v_2 = 0
\end{equation}
for the appearance of the massless graviton on negative-tension brane. Fig. 2 shows
$m^2 \hat{G}_0^{\xi_1, \xi_2}[L, L; m^2 / 2]$ where $\xi_1 = 1/2$ and 
$\xi_2$ is determined from Eq.(\ref{rs1wrongcondi}). Fig. 2 shows again
that there is no massless graviton. 
Although we have not tested all kinds of possibility, our numerical results
strongly suggest that there is no room for the appearance of the
massless graviton in negative-tension brane regardless of $\xi_1$, $\xi_2$, and $L$.

Of course, one can derive a fixed-energy amplitude for 
$\xi_1 = \xi_2 = 0$ case via the coupling constant renormalization in 
principle. However, long expression for $\hat{G}_{RS1}$ seems to make 
the calculation too tedious. So, we do not describe the result of this case
in this paper.

\section{Conclusion}
In this paper we have examined the localized gravity on the brane in RS
brane-world scenario from the singular quantum mechanics. Choosing a single 
copy of $AdS_5$ as a bulk spacetime we have shown that the fixed-energy 
amplitude for RS1 and RS2 are non-trivially dependent on the BCs.

As a result the fixed-energy amplitude for RS2 is dependent on the 
free parameter $\xi$, which parametrize the BC at $y = 0$. Computing the 
fixed-energy amplitude explicitly one can derive the general criterion
(\ref{rs2condi}) for the appearance of the localized massless graviton 
on the brane when $\xi$ is arbitrary but non-zero. When $\xi = 0$, the 
massless graviton is obtained via the coupling constant renormalization.

In RS1 scenario the final expression of the fixed-energy amplitude is 
dependent on the two free parameters $\xi_1$ and $\xi_2$, which parametrize
the various BCs at the end-points of $1d$ box. The appearance of the massless
graviton is numerically tested by examing the pole at $m^2 = 0$. For the 
positive-tension brane our numerical test indicates that there is no 
massless graviton if the length of $1d$ box is finite. However, the 
infinite length of $1d$ box makes the graviton localized on the 
positive-tension brane to be massless, which is effectively identical to the
RS2 scenario. For the negative-tension brane our numerical test shows that 
there is no massless graviton regardless of the length of $1d$ box and 
choice of BCs. 

We can consider the various extension for this paper. Firstly, one may 
include the bending effect of the brane in the computation. In this case,
however, the final expression of the linearized fluctuation does not 
seem to be like Schr\"{o}dinger equation. Thus, we think the method used
in Ref.\cite{garr00} is more convenient than the technique
of singular quantum mechanics to treat the bending effect. One can extend 
the method presented in this paper to the higher-dimensional RS
scenario\cite{leb01,bur02}. If one can find a singular brane solution in
the higher-dimensional case, one can apply the self-adjoint extension or 
a coupling constant renormalization to treat the higher-dimensional
$\delta$-function potential. Of course, it is very interesting if we can find 
a singular solution in six dimension because two-dimensional $\delta$-function
potential has various non-trivial properties such as scale anomaly and 
dimensional transmutation\cite{jack91}. One may extend the present paper to
the moving brane picture\cite{kof02}. But it is unclear for us whether or not
the path-integral solution is in this case analytically obtainable. 

We think the most interesting problem is to understand the reason why there is 
no massless graviton in RS1 scenario. This means that the gauge hierarchy 
problem is not compatible with the massless graviton problem. Thus, it seems
to be important to compromise these two distinct phenomena.

\begin{figure}

\caption{$m$-dependence of $m^2 \hat{G}_{RS1}^{1/2, 1/2}[c, c; m^2/2]$. The 
finite but non-zero at $L = \infty$ indicates that the graviton localized
on the positive-tension brane is massless.} 
\end{figure}

\begin{figure}

\caption{$m$-dependence of $m^2 \hat{G}_{RS1}^{1/2, \xi_2}[L, L; m^2/2]$. 
This figure indicates that there is no localized massless graviton on the
negative-tension brane.}
\end{figure}

\epsfysize=25cm \epsfbox{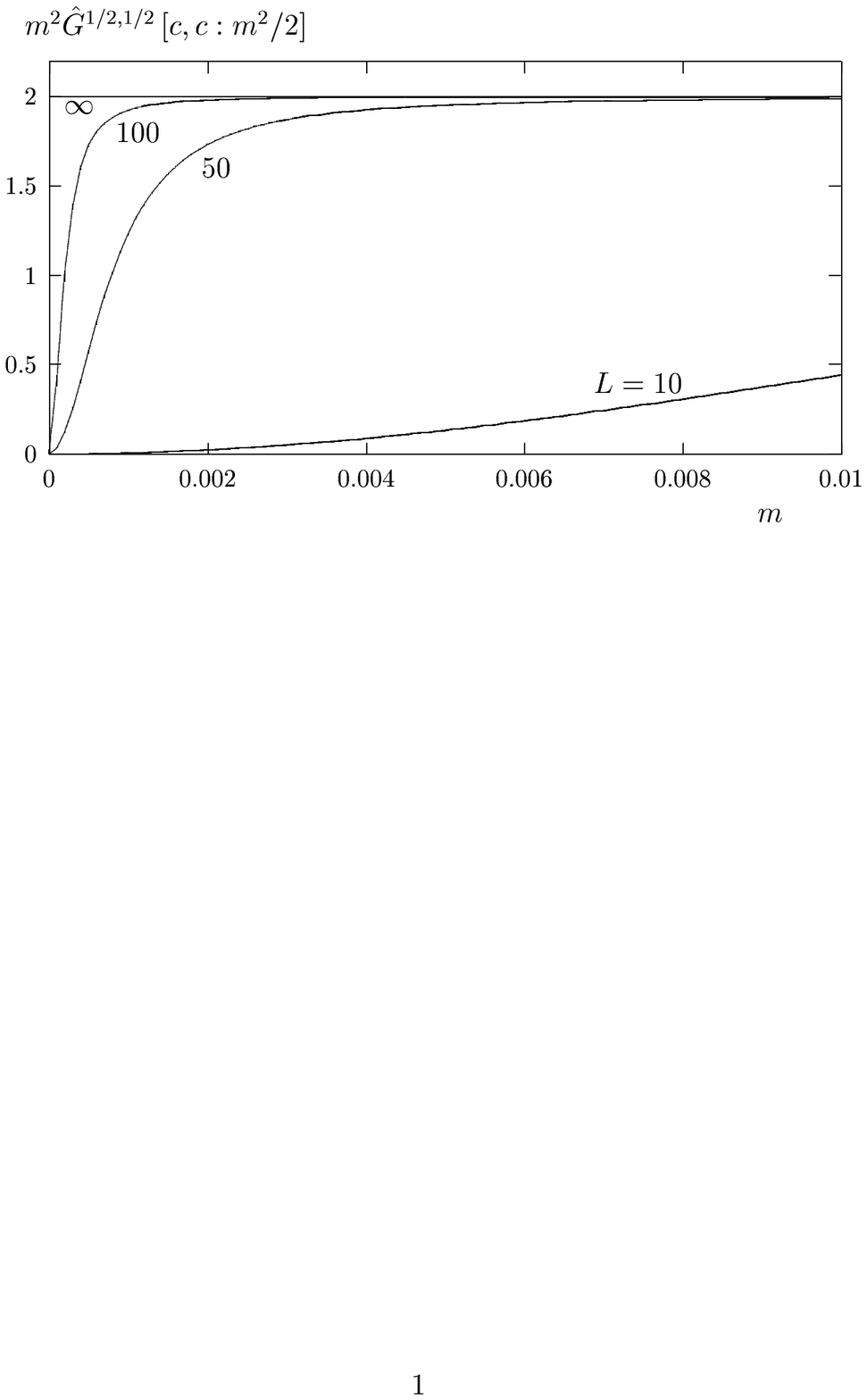}
\epsfysize=25cm \epsfbox{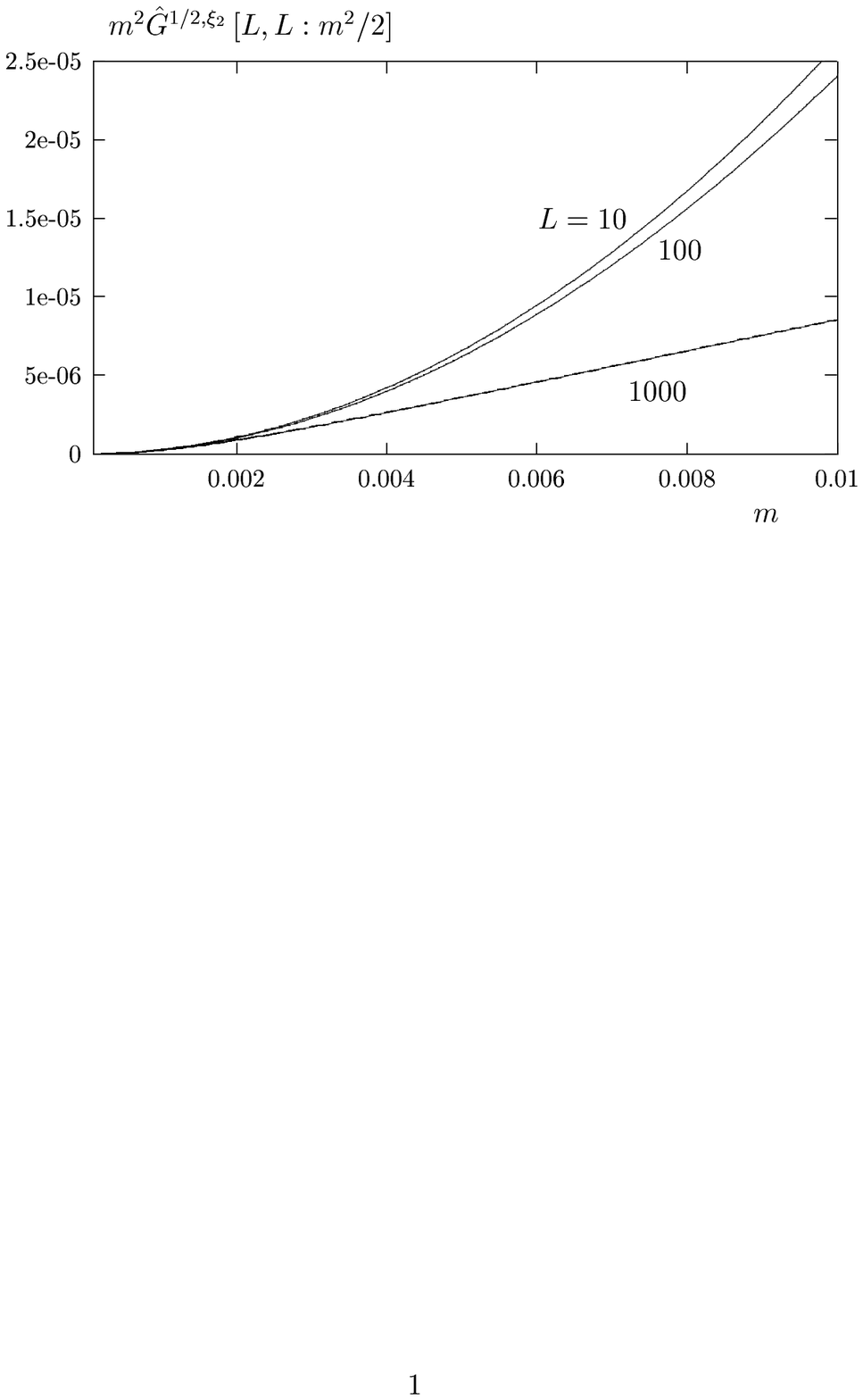}


\begin{thebibliography}{99}
\bibitem{rs99-1} L. Randall and R. Sundrum, Large Mass Hierarchy from a 
Small Extra Dimension, Phys. rev. Lett. {\bf 83} (1999) 3370 [hep-ph/9905221]. 
\bibitem{rs99-2} L. Randall and R. Sundrum, An Alternative to Compactification,
Phys. Rev. Lett. {\bf 83} (1999) 4690 [hep-th/9906064].
\bibitem{garr00} J. Garriga and T. Tanaka, Gravity in the Randall-Sundrum
Brane World, Phys. Rev. Lett. {\bf 84} (2000) 2778 [hep-th/9911055].
\bibitem{duff00} M. J. Duff and J. T. Liu, Complementarity of the Maldacena and
Randall-Sundrum Pictures, Phys. Rev. Lett. {\bf 85} (2000) 2052 [hep-th/0003237].
\bibitem{bine99} P. Binetruy, C. Deffayet and D. Langlois, Non-conventional
Cosmology from a brane universe, Nucl. Phys. {\bf B565} (2000) 269
[hep-th/9905012].
\bibitem{csa99} C. Cs{\'{a}}ki, M. Graesser, C. Kolda, and J. Terning,
Cosmology of One Extra Dimension with Localized Gravity, Phys. Lett.
{\bf B462} (1999) 34 [hep-ph/9906513].
\bibitem{cline99} J. M. Cline, C. Grojean, and G. Servant, Cosmological 
Expansion in the Presence of an Extra Dimension, Phys. Rev. Lett. {\bf 83}
(1999) 4245 [hep-ph/9906523].
\bibitem{kim01} J. E. Kim, B. Kyae, and H. M. Lee, Model for self-tuning
the cosmological constant, Phys. Rev. Lett. {\bf 86} (2001) 4223
[hep-th/0011118].
\bibitem{alex01} S. Alexander, Y. Ling, and L. Smolin, A thermal instability
for positive brane cosmological constant in the Randall-Sundrum cosmologies
[hep-th/0106097].
\bibitem{cham00} A. Chamblin, S. W. Hawking, and H. S. Reall, 
Brane-World black holes, Phys. Rev. {\bf D 61} (2000) 065007 [hep-th/9909205].
\bibitem{emp00} R. Emparan, G. T. Horowitz, and R. C. Myers, Exact Description
of Black Holes on Branes, JHEP {\bf 0001} (2000) 007 [hep-th/9911043].
\bibitem{gidd00} S. B. Giddings, E. Katz, and L. Randall, Linearized Gravity
in Brane Backgrounds, JHEP {\bf 0003} (2000) 023 [hep-th/0002091].
\bibitem{mal98} J. Maldacena, The large-N limit of superconformal field
theories and supergravity, Adv. Theor. Math. Phys. {\bf 2} (1998) 231
[hep-th/9711200].
\bibitem{park01-1} D. K. Park, H. S. Kim, Y. G. Miao, and 
H. J. W. M\"{u}ller-Kirsten, Randall-Sundrum Scenario at Nonzero Temperature,
Phys. Lett. {\bf B 519} (2001) 159
[hep-th/0107156].
\bibitem{park01-2} D. K. Park, H. S. Kim, and S. Tamaryan, Nonvanishing
Cosmological Constant of Flat Universe in Brane-World Scenario, to appear
in Phys. Lett. B [hep-th/0111081].
\bibitem{fey65} R. P. Feynman and A. R. Hibbs, {\it Quantum Mechanics and
Path Integrals} (McGraw-Hill, New York, 1965).
\bibitem{schul81} L. S. Schulman, {\it Techniques and Applications of Path 
Integrals} (Wiley, New York, 1981).
\bibitem{schul86} L. S. Schulman, in {\it Path Integrals from mev to MeV}, 
edited by M. C. Gutzwiller, A. Inomata, J. R. Klauder, and L. Streit 
(World Scientific, Singapore, 1986).
\bibitem{park95} D. K. Park, Green's-function approach to two- and 
three-dimensional delta-function potential and application to the spin-1/2 
Aharonov-Bohm problem, J. Math. Phys. {\bf 36} (1995) 5453 [hep-th/9405020].
\bibitem{jack91} R. Jackiw, in {\it M. A. B\'{e}g Memorial Volume}, edited by 
A. Ali and P. Hoodbhoy (World Scientific, Singapore, 1991).
\bibitem{capri85} A. Z. Capri, {\it Nonrelativistic Quantum Mechanics}
(Benjamin/Cummings, Menlo Park, 1985).
\bibitem{albev88} S. Albeverio, F. Gesztesy, R. Hoegh-Krohn, and H. Holden,
{\it Solvable Models in Quantum Mechanics} (Springer, Berlin, 1988).
\bibitem{gros93} C. Grosche, $\delta$-function perturbations and boundary
problems by path integration, Ann. Physik {\bf 2} (1993) 557 [hep-th/9302055].
\bibitem{gros95} C. Grosche, $\delta^{\prime}$-Function Perturbations and 
Neumann Boundary-Conditions by Path Integration, J. Phys. {\bf A28} (1995) L99
[hep-th/9402110].
\bibitem{park96} D. K. Park, Proper incorporation of self-adjoint extension
method to Green's function formalism: one-dimensional 
$\delta^{\prime}$-function potential case, J. Phys. {\bf A29} (1996)
6407 [hep-th/9512097].
\bibitem{park98} D. K. Park and S. K. Yoo, Propagators for spinless and 
spin-1/2 Aharonov-Bohm-Coulomb Systems, Ann Phys. {\bf 263} (1998)
295 [hep-th/9707024].
\bibitem{park01-3} D. K. Park and S. Tamaryan, Compromise of Localized Graviton
with a Small Cosmological Constant in Randall-Sundrum Scenario, 
Phys. Lett. {\bf B532} (2002) 305 [hep-th/0108068].
\bibitem{gros98} C. Grosche and F. Steiner, {\it Handbook of Feynman Path 
Integrals} (Springer, Berlin, 1998).
\bibitem{leb01} F. Leblond, R. C. Myers, and D. J. Winters, Brane World
sum rule and $AdS$ Soliton [hep-th/0107034].
\bibitem{bur02} C. P. Burgess, J. M. Cline, and N. R. Constable, and 
H. Firouzjahi, Dynamical Stability of Six-Dimensional Warped Brane-Worlds,
JHEP {\bf 0201} (2002) 014 [hep-th/0112047].
\bibitem{kof02} G. Kofinas, New Perspectives on Moving Domain Wall in 
$(A)dS_5$ space, Nucl. Phys. {\bf B 622} (2002) 347 [hep-th/0103045].
\end{thebibliography}
\end{document}